\documentclass[aps,prl,twocolumn,showpacs]{revtex4}
\usepackage{graphicx}
\usepackage{float}

\begin{document}


\title{Combined constraints on Majorana masses \\ from neutrinoless double beta decay experiments}



\author{Steven D. Biller  }

\email{steven.biller@physics.ox.ac.uk}
\affiliation{Department of Physics, University of Oxford, Oxford OX1 3RH, United Kingdom} 

\date{\today}

\begin{abstract}
Combined bounds on the Majorana neutrino mass for light and heavy neutrino exchange mechanisms are derived from current neutrinoless double beta decay ($0\nu\beta\beta$) search results for a variety of nuclear matrix element (NME) models. The approach requires self-consistency of a given model to predict NMEs across different isotopes. The derived bounds are notably stronger than those from any single experiment and show less model-to-model variation,  highlighting the advantages of using multiple isotopes in such searches. Projections indicate that the combination of near-term experiments should be able to probe well into the inverted neutrino mass hierarchy region. A method to visually represent $0\nu\beta\beta$ experimental results is also suggested to more transparently compare across different isotopes and explicitly track model dependencies.
\end{abstract}


\maketitle


The origin of neutrino mass is one of the central puzzles in particle physics today. It is intimately connected to the question of whether neutrinos are Dirac 
or Majorana particles, 
with fundamental implications for both particle physics and cosmology. Owing to the small scale of neutrino masses, the search for neutrinoless double beta decay ($0\nu \beta \beta$) is the only known experimental approach that can be practically used to address this question. 

However, the translation of experimental half-life bounds from different $0\nu \beta \beta$ isotopes into Majorana mass limits is complicated by the fact that the nuclear matrix elements (NMEs) for the transition currently have a high degree of uncertainty. There are a variety of different theoretical approaches and parameter choices ({\em e.g.} values of coupling constants and correlation functions) that result in predictions varying by a factor of two or more, often with different relative values between isotopes \cite{NMEs}. One particularly important parameter choice relates to the question of a potential ``quenching'' of the effective axial vector coupling constant, $g_A$, based on discrepancies observed in the rates of single beta decays relative to calculations \cite{quenching}. Recent {\it ab initio} calculations of some beta decay isotopes suggest that the equivalent suppression of the NMEs naturally arises as the result of meson currents and other higher order nucleon interactions \cite{abinitio}. Such interactions are expected to play less of a role for the higher momentum transfers related to $0\nu \beta \beta$, but it is widely believed that some suppression will result, though the details remain uncertain. This situation significantly complicates the extraction of robust bounds on Majorana neutrino masses as well as the comparison and combination of experimental results between different isotopes.

The approach taken here is to explicitly treat each NME formalism and set of parameter choices as a separate, self-consistent model that makes linked predictions between different $0\nu \beta \beta$ isotopes. As such, different experimental results for a given model can then be compared and combined to yield improved mass limits. 
This approach is similar to that of \cite{Guzowski}, but here explicitly setting bounds on the model space for light and heavy neutrino exchange mechanisms and making use of likelihood functions based on recent experimental results and NME calculations. In what follows, NME calculations for an unquenched value of $g_A=1.27$ will be used throughout, however the first order dependence of $g_A$ on the half-life (which actually resides in the isotope-specific phase space factor) will be factored out for visibility and to retain model-independence of the isotopic correction.

\section{I. Experimental Likelihood Functions}

Four experimental results will be considered in this analysis: CUORE \cite{CUORE}, EXO-200 \cite{EXO-200}, GERDA Phase II \cite{GERDA} and KamLAND-Zen \cite{KZ1,KZ2}. The combination of results will make use of the corresponding likelihood functions.

For CUORE, the likelihood function was directly provided in their paper in terms of a posterior probability with uniform prior as a function of assumed $0\nu\beta\beta$ decay rate, which has been extracted and parameterized by a polynomial in log(likelihood) over the observable range. 

The GERDA Phase II experiment observed zero counts in the region of interest, making the likelihood function trivially a Poisson distribution with $n=0$, which is just the exponential $e^{-(\mu_S + \mu_B)}$, where $\mu_S$ and $\mu_B$ are the expected mean number of signal and background events, respectively. However, for relative likelihoods, $\mu_B$ can be ignored, leaving just $e^{-\mu_S}$.

EXO-200 results are summarized in their 2019 paper, which gives the numbers of events observed within a $\pm$2$\sigma$ window for phases 1 and 2, along with the expected background levels and their uncertainties. A likelihood function was therefore constructed based on a Poisson distribution for each phase, taking into account slightly different $0\nu\beta\beta$ detection efficiencies, and convolving these with a Gaussian for the background uncertainties. A combined likelihood was then constructed for a given signal hypothesis, accounting for the exposures of each dataset.

The derived likelihood function for KamLAND-Zen was based on their published observations for the numbers of events seen in each 50~keV energy bin near the signal region compared with their best fit background model. The potential signal fraction in each bin was calculated from a normal distribution centered on the $^{136}$Xe endpoint, with a width commensurate with their stated energy resolution. For simplicity, a Poisson probability distribution was used for each bin assuming that the background model was perfectly constrained by data outside the region of interest. This errs on the side of providing slightly tighter constraints. Separate likelihood constructions were made for Phase I \cite{KZ1}, Phase II Period~1 and Phase II Period 2 \cite{KZ2}. Once more, a combined likelihood was then constructed for a given signal hypothesis, taking into account the exposures of each dataset.

Figure 1 shows the resulting values of $-2\log L_R$ versus the assumed $0\nu\beta\beta$ signal rate, where $L_R$ is the likelihood ratio with respect to the maximum for each experiment.

\begin{center}
\begin{figure}
\includegraphics[width=67mm]{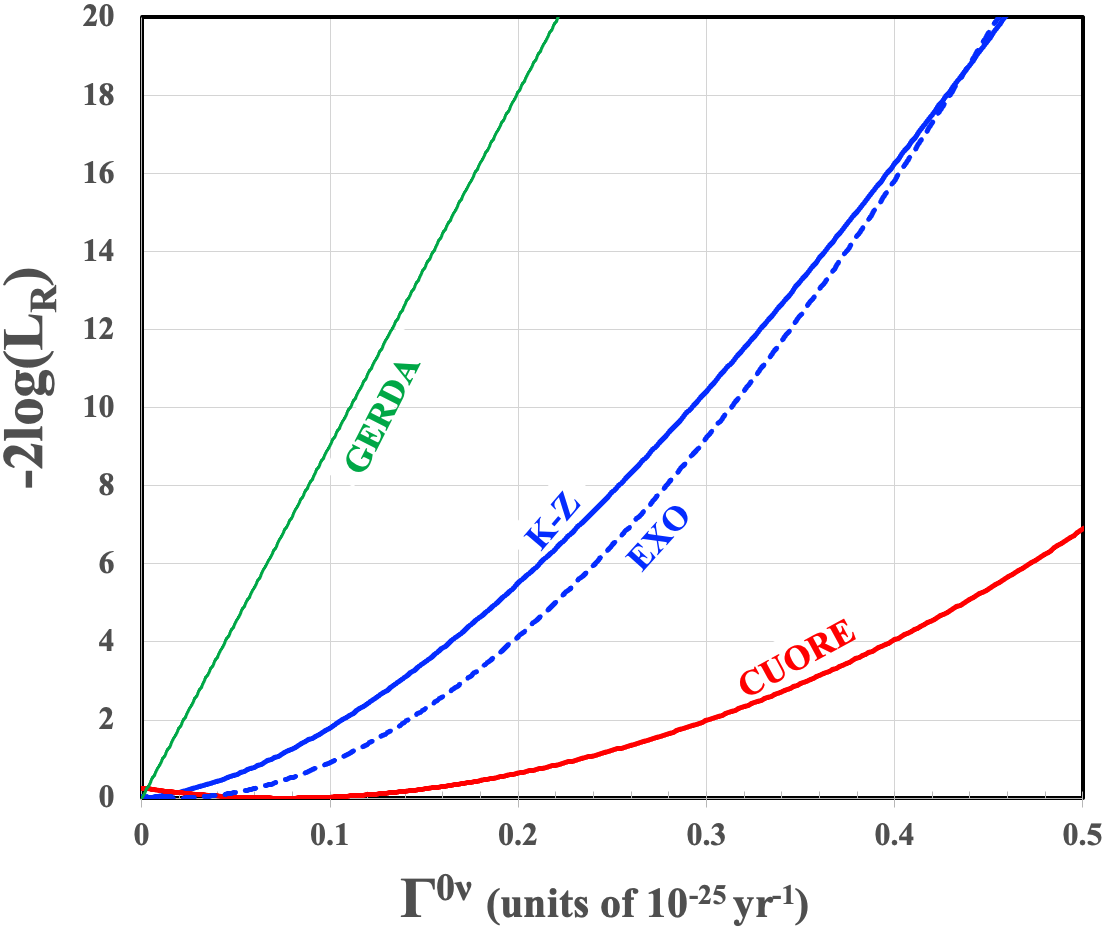}
\caption{Derived/parameterized values of $-2\log(L_R)$ for the various experiments as a function of assumed $0\nu\beta\beta$ decay rate.}
\end{figure}
\end{center}

\section{II. Construction of Half-Life Bounds}

To specifically constrain the phase space of possible models, 90\% CI Bayesian bounds with a prior that is uniform in the (positive) counting rate will be used here. For counting experiments searching for rare events, this tends to provide conservative limits that are more robust to statistical background fluctuations than many frequentist constructions while, conveniently, also yielding a relatively close correspondence with numerical values characteristic of frequentist coverage \cite{Confidence}. For $0\nu\beta\beta$, such a choice corresponds to a prior that is uniform in $m_{\beta\beta}^2$ for the light neutrino exchange (LNE) mechanism. This, indeed, can be seen to yield more conservative upper bounds than, for example, a prior that is uniform in either $m_{\beta\beta}$ or $\log m_{\beta\beta}$. For the heavy neutrino exchange (HNE) mechanism, where one sets lower bounds on the mass scale of the heavy neutrino, a prior uniform in rate corresponds to a prior uniform in $M_{\beta\beta}^{-2}$, which is, again, a more conservative prior choice for a lower bound than one that is uniform in either $M_{\beta\beta}$ or $\log M_{\beta\beta}$.

90\% CI upper bounds on the $0\nu\beta\beta$ decay rate, $\Gamma^{0\nu}$, have therefore been derived by integrating the posterior probability (which, for a uniform prior, is equivalent to the normalized $L_R$ values inferred from Fig.~1) as a function of the event rate from zero until the point where 90\% of the distribution is retained. The 90\% CI upper bound on the half-life is then $\log(2)/\Gamma^{0\nu}_{90\%}$. The Bayesian bounds thus produced exactly match those published by CUORE and GERDA. 
The derived half-life bounds for EXO-200 are slightly more restrictive in appearance than their published frequentist limit ($4.3\times10^{25}$ yr versus $3.5\times10^{25}$ yr), which might be partly due to a more detailed treatment within their analysis window. On the other hand, the numerical value of the KamLAND-Zen bound is roughly a factor of $\sim$2 lower than their published frequentist limit ($4.9\times10^{25}$ yr versus $1.07\times10^{26}$ yr). In both cases, the Bayesian bounds are closer to the expected median experimental sensitivities.

\section{III. Model-By-Model Constraints}

The $0\nu\beta\beta$ half-life, $T_{1/2}^{0\nu}$, can be related to the effective Majorana neutrino masses (as defined by the PMNS mixing matrix $U$) according to \cite{Simkovic}:
\begin{equation}
\left(T_{1/2}^{0\nu}\right)^{-1} = \frac{G^{'0\nu}}{g_A^4}  \left| \mathcal{M}_{L}^{0\nu}\left(\frac{m_{\beta\beta}}{m_e}\right) + \mathcal{M}_{H}^{0\nu} \left(\frac{m_p}{M_{\beta\beta}}\right) \right|^2
\end{equation}
\noindent where $\mathcal{M}_{L}^{0\nu}$ is the NME for the LNE transition;  $\mathcal{M}_{H}^{0\nu}$ is the NME for the HNE transition; $m_{\beta\beta}$ is the effective Majorana light neutrino mass $\left(\equiv \left|\sum_l U_{el}^2 m_l \right| \right)$; $M_{\beta\beta}$ is the effective heavy neutrino mass $\left(\equiv \left|\sum_l U_{el}^2 /M_l \right| ^{-1}\right)$;  and the factorized form of the isotopic phase space factor, $G^{'0\nu} = g_A^4 G^{0\nu}$, is used to separate the dependence on the axial-vector coupling, $g_A$.
For simplicity, it will be assumed that we are in a region where one of these two terms in dominant, allowing bounds to be separated placed on LNE and HNE mechanisms.

For a given value of NME, the likelihoods of Fig.~1 can therefore be translated into functions of $m_{\beta\beta}$ or $M_{\beta\beta}$. The likelihoods can then be combined and used to yield 90\% CI bounds on the mass scales for LNE and HNE, as previously described, by integrating the posterior probability as a function of $m_{\beta\beta}^2$ or $M_{\beta\beta}^{-2}$, respectively. Table~I shows the NME values predicted by a variety of different models, with details of model assumptions provided in the corresponding references. While not exhaustive, this list typifies the span of variation between model predictions. Figure~2 shows examples of the interplay between the experimental likelihoods for four of those NME models in the case of LNE. Derived upper bounds are indicated by filled circles both for the combined likelihood as well as for those of individual experiments. Table~II summarizes the derived combined constraints on both LNE and HNE  for each NME model.

\begin{figure}
\includegraphics[width=80mm]{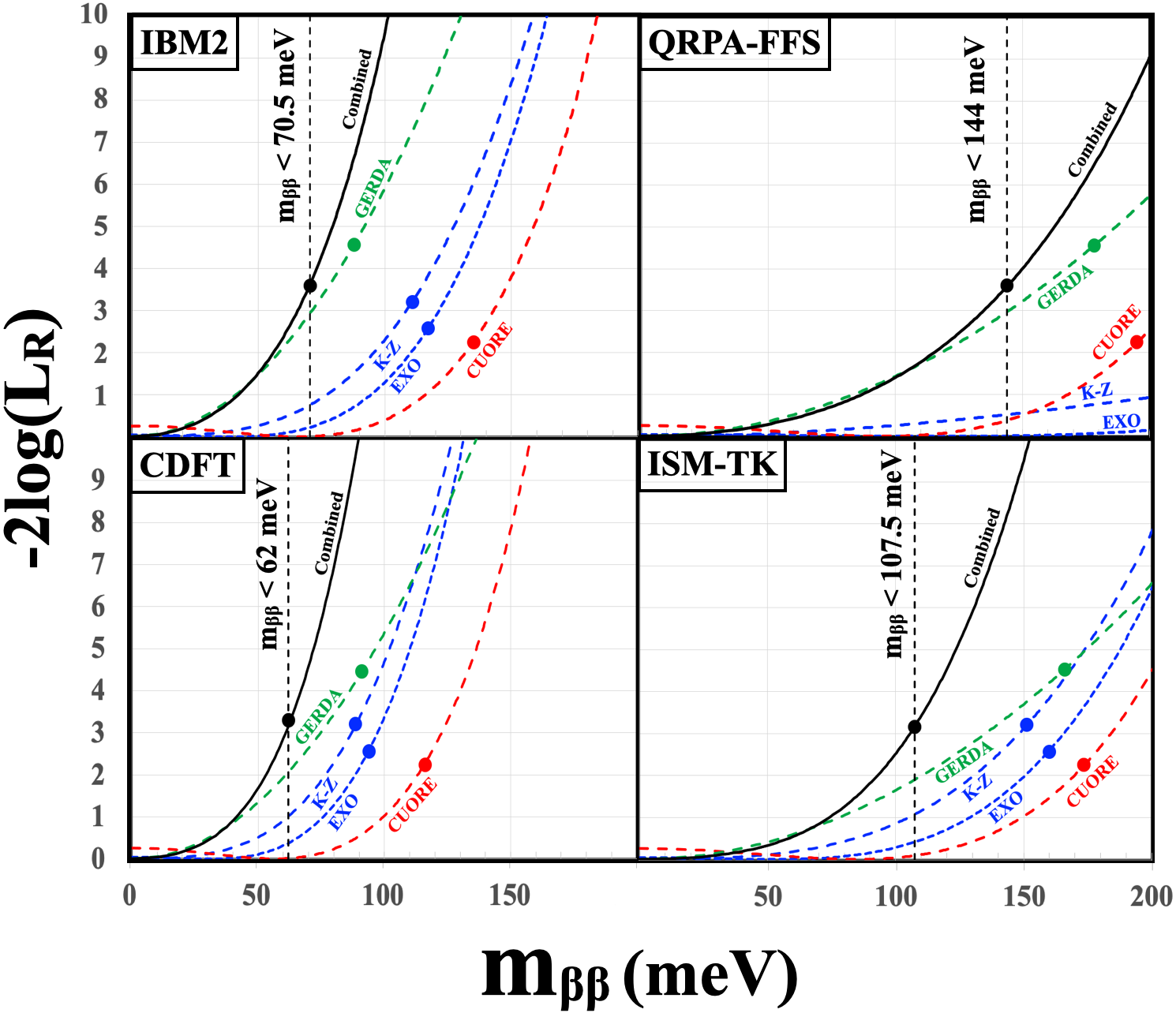}
\caption{$-2\log(L_R)$ as a function of $m_{\beta\beta}$ for four of the NME models considered for the LNE mechanism. Curves for individual experiments are shown as well as the combined likelihood. Solid circles indicate 90\% CI upper bounds for each likelihood curve for priors uniform in $m_{\beta\beta}^2$.}
\end{figure}

\begin{table}
\begin{tabular}{|c|l|l|l|} \hline
NME Model & \multicolumn{3}{|c|}{LNE (HNE) Matrix Elements} \\
 ($g_A$=1.27)& \multicolumn{1}{|c|}{$^{76}$Ge} &  \multicolumn{1}{|c|}{$^{130}$Te} &  \multicolumn{1}{|c|}{$^{136}$Xe} \\ \hline
IBM2 \cite{IBM2} & 6.34 (181.6) & 4.2 (126.8) & 3.4 (99.2)\\
CDFT \cite{CDFT} & 6.04 (209.1) & 4.89 (193.8) & 4.24 (166.3) \\
QRPA-FFS \cite{QRPA-FFS} & 3.12 (187.3)  & 2.9 (191.4) & 1.11 (66.9) \\
QRPA-JY \cite{QRPA-JY} & 5.26 (401.3)& 4.0 (338.3) & 2.91 (186.3)  \\
QRPA-TU \cite{QRPA-TU1, QRPA-TU2} & 5.16 (287) & 2.89 (264) & 2.18 (152)  \\
ISM-TK \cite{ISM-TK} & 2.89 (130) & 2.76 (146) & 2.28 (116) \\
QRPA-NC \cite{QRPA-NC} & 5.09 & 1.37 & 1.55\\
ISM-INFN \cite{ISM-INFN} & 3.34 & 3.26  & 2.49  \\ 
CGM \cite{CGM} & 5.518 & 6.366  & 4.755  \\ 
\hline
\end{tabular}
\caption{Matrix elements for various NME models.}
\end{table}

\begin{table}
\begin{tabular}{|c|c|c|} \hline
& Comb. 90\% CI & Comb. 90\% CI  \\ 
NME Model & Upper Bound& Lower Bound \\ 
 ($g_A$=1.27)& on $m_{\beta\beta}$ (meV) & on $M_{\beta\beta}$ (GeV)\\ \hline
IBM2 \cite{IBM2} & 70.5 & $7.6\times 10^7$\\
CDFT \cite{CDFT} & 62 & $1.1\times 10^8$\\
QRPA-FFS \cite{QRPA-FFS} & 144 & $7.9\times 10^7$\\
QRPA-JY \cite{QRPA-JY} & 82 & $1.6\times 10^8$\\
QRPA-TU \cite{QRPA-TU1, QRPA-TU2} & 92.5 & $1.25\times 10^8$\\
ISM-TK \cite{ISM-TK} & 107.5 & $7.6\times 10^7$\\
QRPA-NC \cite{QRPA-NC} & 100 & \\
ISM-INFN \cite{ISM-INFN} & 120 &\\ 
CGM \cite{CGM} & 57 &\\ 
\hline
\end{tabular}
\caption{Derived 90\% CI combined bounds for LNE and HNE mass scales for various NME models.}
\end{table}

\begin{figure}
\includegraphics[width=83mm]{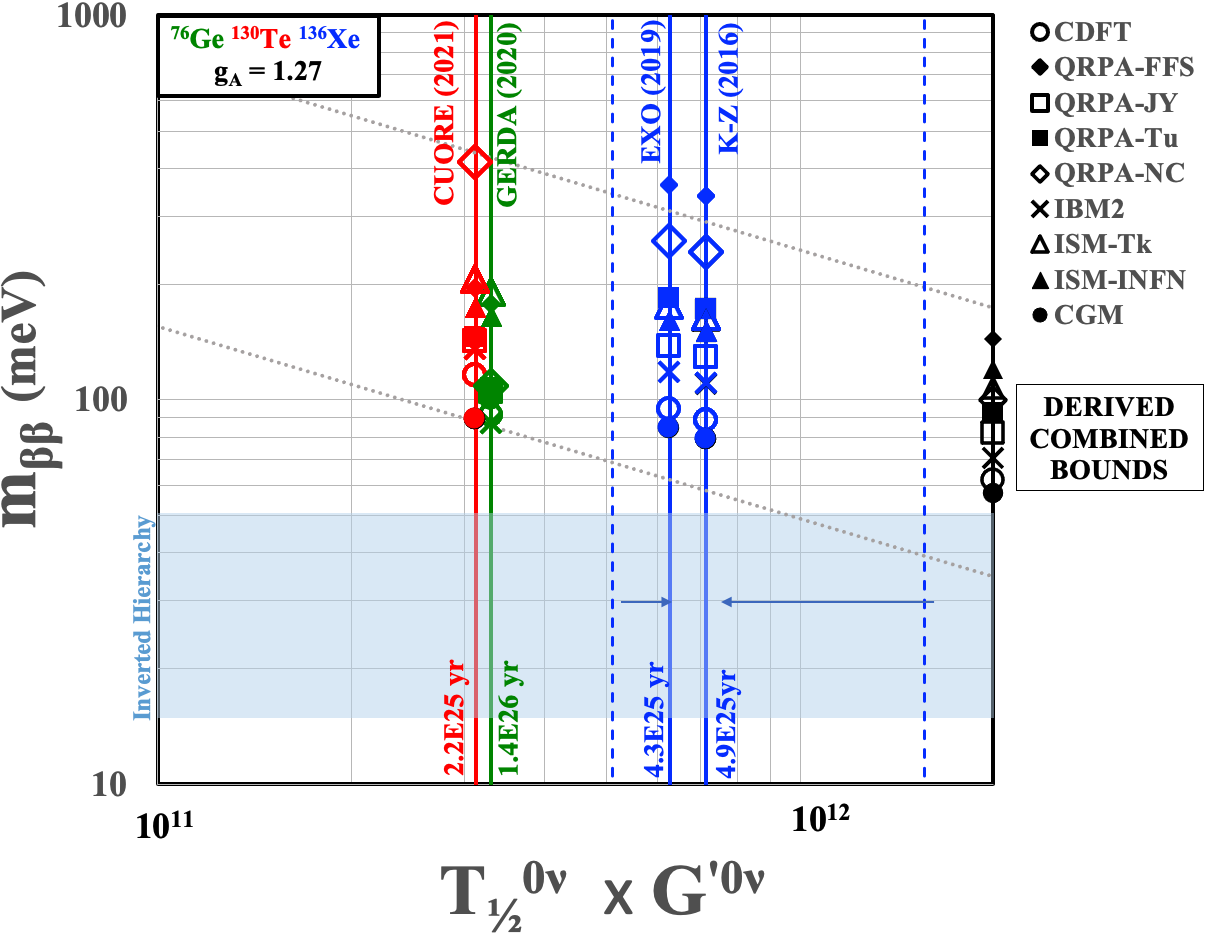}
\caption{Comparison plot of 90\% CI upper bounds for the LNE mechanism. Individual and combined constraints on $m_{\beta\beta}$ for different NME models are indicated by symbols.}
\end{figure}

\begin{figure}
\includegraphics[width=83mm]{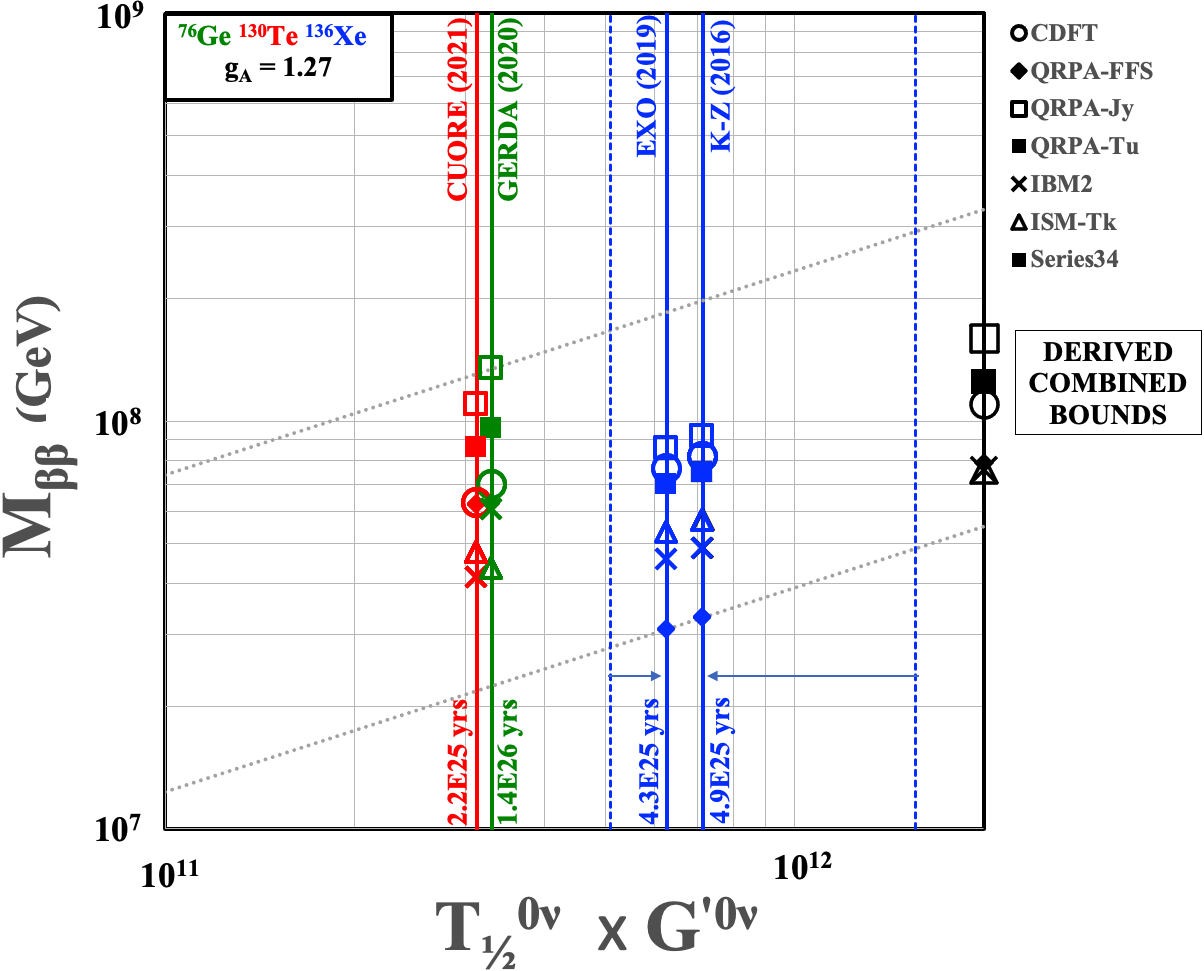}
\caption{Comparison plot of  90\% CI lower bounds for the HNE mechanism. Individual and combined constraints on $M_{\nu h}$ for different NME models are indicated by symbols.}
\end{figure}

\section{IV. Visual Representation for Transparent Comparisons}

In order to make the nature of the model dependencies as transparent as possible when comparing experimental sensitivities, and to facilitate the relevant comparison across different isotopes, Figures 3 and 4 (for LNE and HNE, respectively) plot the product of $T_{1/2}^{0\nu} \times G^{'0\nu}$ for the different experimental results along the x-axis. This unitless ``TG sensitivity'' provides a useful model-independent, first order isotope correction to indicate the rough physics reach of a given observation before the uncertain NME values are taken into account. 

A given experimental bound results in a vertical line at the corresponding TG value, with the half-life bound also given at the base of each line. The two dashed vertical lines are here used to indicate the difference between the Bayesian bounds and published frequentist values for EXO-200 and KamLAND-Zen. Different matrix element models are represented by the different symbols indicated, each allowing a model-dependent translation to an effective neutrino mass on the y-axis. Different isotopes are color coded and the diagonal dotted black lines indicate how the mass sensitivity scales with TG sensitivity. The region approximately corresponding to the inverted neutrino mass hierarchy in Fig. 3 is indicated by the horizontal blue band (which also includes part of the normal mass hierarchy for degenerate neutrinos). The combined bounds derived in this work are indicated on the right side of each plot. These bounds are notably more restrictive than those from any single experiment.

\section{V. Projected Combined Bounds for Near-Term Experiments}

A similar approach has been taken to project the physics reach for the combination of running and near-term experiments: CUORE, KamLAND-Zen 800, LEGEND-200 and SNO+ I. Five years of live time with the full experiment was assumed for each. 

For CUORE and LEGEND-200, a single-bin Poisson model spanning $\pm1.5\sigma$ about the endpoint was used for the likelihood, assuming similar signal efficiencies as current versions of the two technologies. In the case of CUORE, a background index of $1.38\times 10^{-2}$ counts/keV-kg($TeO_2$)-yr was used \cite{CUORE} and, for LEGEND-200, a value of $2\times10^{-4}$ counts/keV-kg($^{76}Ge$)-yr was assumed \cite{LEGEND}.

For KamLAND-Zen and SNO+, a Poisson-based likelihood with multiple energy bins in the region of interest was used. Background contributions from $^8B$ solar neutrinos and $2\nu\beta\beta$ ``spill-over" due to energy resolution were directly calculated. In the case of SNO+, additional internal and external backgrounds were added to produce a model consistent with reference \cite{Grant}. For KamLAND-Zen, it was assumed that U/Th backgrounds could be reduced to negligible levels by improvements to analysis and electronics and that the fiducial volume could be extended to a radius of 1.65m so as to achieve their sensitivity goal of $T_{1/2} > 5\times10^{26}$~yrs \cite{Grant}. 

The  resulting 90\% CI sensitivities assuming no observed excess for both individual and combined likelihoods (including current measurements) are shown in Fig. 5. Again, projected combined bounds are significantly more restrictive than those from any single measurement and suggests that the combination of near-term experiments will be able to probe well into the region corresponding to the inverted neutrino mass hierarchy.

\begin{figure}
\includegraphics[width=85mm]{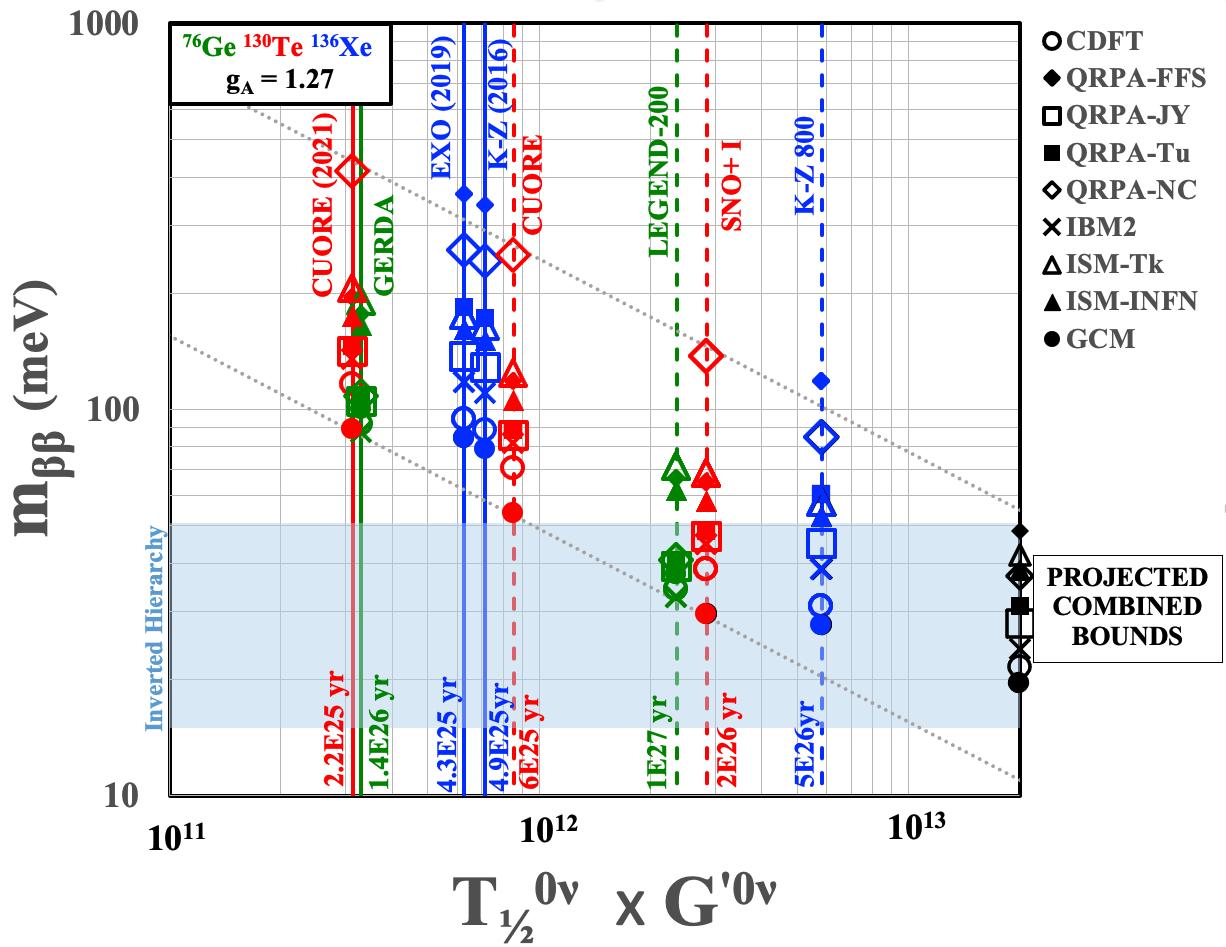}
\caption{Current (solid) and projected near-term (dashed) sensitivities of $0\nu\beta\beta$ experiments for the LNE mechanism. Individual and combined constraints on $m_{\beta\beta}$ for different NME formalisms are indicated by symbols.}
\end{figure}

\section{VI. Conclusion}
Combined constraints on Majorana masses have been derived based on the results of recent $0\nu\beta\beta$ experiments by demanding self-consistency of any given NME model (defined as the combination of formalism and parameter choice) in its predictions across different isotopes. Parameterized or derived likelihood functions for results from CUORE, EXO-200, GERDA and KamLAND-Zen were used to arrive at 90\% CI bounds for LNE and HNE mechanisms using a conservative choice of prior. The derived combined constraints are notably more restrictive than those from any one experiment and also have less variation across different NME models, highlighting the complementarity of these different approaches and the advantages of using different isotopes in the search for $0\nu\beta\beta$. While uncertainties on NME values remain, projections for near-term experiments suggest that their combination will be able to probe well into the IH region for the NME models considered here. A method to visually display $0\nu\beta\beta$ results has also been suggested to allow for the transparent comparison of sensitivities and model-dependencies across different isotopes.

\begin{acknowledgements}
The author would like to thank Frank Deppisch and Josh Klein for helpful discussions. This work is supported by the Science and Technology Facilities Council of the United Kingdom.
\end{acknowledgements}


\begin{thebibliography}{99}
\bibitem{NMEs}  J. Engel and J. Menendez, Rep. Prog. Phys. {\bf 80}, 046301 (2017) 
\bibitem{quenching} H. Ejiri, Front. Phys., (2019)
\bibitem{abinitio} P. Gysbers {\em et al.}, Nat. Phy. {\bf 15}  (2019)
\bibitem{Guzowski} P. Guzowski {\em et al.}, Phys. Rev. D {\bf 92}, 012002 (2015)
\bibitem{CUORE} D. Q. Adams {\em et al.}, arXiv:2104.06906 
\bibitem{EXO-200} G. Anton {\em et al.}, Phys. Rev. Lett. {\bf 123}, 161802 (2019)
\bibitem{GERDA} M. Agostini {\em et al.}, Phys. Rev. Lett. {\bf 125}, 252502 (2020)
\bibitem{KZ1} A. Gando {et al.}, Phys. Rev. Lett. {\bf 110}, 062502 (2013)
\bibitem{KZ2} A. Gando {\em et al.}, Phys. Rev. Lett. {\bf 117}, 082503 (2016)
\bibitem{Confidence} S. Biller and S. Oser, NIM {\bf A774} (2015)
\bibitem{Simkovic} F. Šimkovic, G. Pantis, J. D. Vergados, and A. Faessler
Phys. Rev. C {\bf 60}, 055502 (1999)
\bibitem{IBM2} F. Deppisch {\em et al.}, Phys. Rev. D {\bf 102}, 095016 (2020)
\bibitem{CDFT} L. S. Song, J. M. Yao, P. Ring, and J. Meng, Phys. Rev. C {\bf 95}, 024305 (2017)
\bibitem{QRPA-FFS} Dong-Liang Fang, Amand Faessler, and Fedor Šimkovic, Phys. Rev. C {\bf 97}, 045503 (2018)
\bibitem{QRPA-JY} J. Hyvarinen and J. Suhonen, Phys. Rev. C {\bf 91}, 024613 (2015)
\bibitem{QRPA-TU1} F. Šimkovic, V. Rodin, A. Faessler and P. Vogel, Phys.Rev. C {\bf 87}, 045501 (2013)
\bibitem{QRPA-TU2} A. Faessler, M. Gonzalez, S. Kovalenko and F. Šimkovic, Phys. Rev. D {\bf 90}, 096010 (2014)
\bibitem{ISM-TK} J. Menendez, J. Phys. G: {\bf 45},  014003 (2018 )
\bibitem{QRPA-NC} M. T. Mustonen and J. Engel, Phys. Rev. C {\bf 87}, 064302 (2013)
\bibitem{ISM-INFN} L. Coraggio {\em et al.}, Phys. Rev. C 101, 044315 (2020)
\bibitem{CGM} N. Vaquero, T. Rodriguez and J. Egido, Phys. Rev. Lett. {\bf 111}, 14501 (2013)
\bibitem{LEGEND} F. Avignone and S. Elliott, Front. Phys. {\bf 7,} 6 (2019) 
\bibitem{Grant} C. Grant, XXIX International Conference on Neutrino Physics, Fermilab (2020)


\end{thebibliography}
\end{document}